# Interface-Sensitive Raman Microspectroscopy of Water via Confinement with a Multimodal Miniature Surface Forces Apparatus

Hilton B. de Aguiar,[1,*] Joshua D. McGraw,[1,2] Stephen H. Donaldson Jr.[1,*]

[1]Département de Physique, Ecole Normale Supérieure/PSL Research University, CNRS, 24 rue Lhomond, 75005 Paris, France

[2]Gulliver CNRS UMR 7083, PSL Research University, ESPCI Paris, 10 rue Vauquelin, 75005 Paris, France

*Corresponding authors: h.aguiar@phys.ens.fr; steve.donaldson@phys.ens.fr

**Abstract**

Modern interfacial science is increasingly multi-disciplinary. Unique insight into interfacial interactions requires new multimodal techniques for interrogating surfaces with simultaneous complementary physical and chemical measurements. We describe here the design and testing of a microscope that incorporates a miniature Surface Forces Apparatus (μSFA) in sphere vs. flat mode for force-distance measurements, while simultaneously acquiring Raman spectra of the confined zone. The microscope uses a simple optical setup that isolates independent optical paths for (*i*) the illumination and imaging of Newton's Rings and (*ii*) Raman-mode excitation and efficient signal collection. We benchmark the methodology by examining Teflon thin films in asymmetric (Teflon-water-glass) and symmetric (Teflon-water-Teflon) configurations. Water is observed near the Teflon-glass interface with nanometer-scale sensitivity in both the distance and Raman signals. We perform chemically-resolved, label-free imaging of confined contact regions between Teflon and glass surfaces immersed in water. Remarkably, we estimate that the combined approach enables vibrational spectroscopy with single water monolayer sensitivity within minutes. Altogether, the Raman-μSFA allows exploration of molecular confinement between surfaces with chemical selectivity and correlation with interaction forces.

**Introduction**

Over the past five decades, the surface forces apparatus (SFA) has become a standard tool for measuring interaction forces between extended surfaces. It has been used to gain quantitative insights into so-called DLVO forces[1] (*i.e.*, the combination of electrostatic double-layer and van der Waals forces), solvation forces,[2] polymer brush interactions,[3,4] friction and lubrication,[5,6] hydrophobic interactions,[7,8] biomimetic adhesion,[9,10] and protein interactions.[11–13] Now, to examine a broader range of systems and gain deeper physicochemical insights, it is advantageous to couple the SFA with additional measurement probes *in situ*.

The SFA has previously been combined with X-Ray diffraction,[14,15] electrochemistry,[16–19] fluorescence microscopy,[20–22] and Raman spectroscopy.[23,24] These efforts are often limited by the SFA itself, which is a bulky instrument and requires long working distance (WD) microscope objectives to reach the focal plane of the SFA contact zone. High resolution microscopy requires high numerical aperture (NA) objective lenses that usually have a small WD, and is therefore especially difficult in a standard SFA. The same limitation applies for spectroscopic methods such as spontaneous Raman scattering, which generally has a weak signal and requires high NA to gather enough light for a measureable signal. Indeed, only one example of combined SFA-Raman spectroscopy can be found in the literature, in which Granick and coworkers found different degrees of polydimethylsiloxane chain ordering as a function of confinement and shear in a custom built confocal Raman-SFA.[23,24] Other than SFA, a tribometer was coupled with total internal reflection Raman to observe correlations between film thickness, frictional forces, and Raman signal of confined liquids, particles, and oils.[25–27] Recently, crystalline ordering of pentadecane was observed at the contact line between swollen polydimethylsiloxane and sapphire by combining a tribometer with sum frequency generation (SFG).[28] In general, previous work has not exploited microspectroscopy, leaving open interesting questions related to confinement-dependent morphology changes.[20]

To increase the applicability of SFA for such cross-disciplinary multi-modality, Israelachvili and coworkers recently developed a miniature version of the SFA, called the µSFA, which is designed to be mounted and operated on standard inverted optical microscopes. We and others showed that the µSFA is easily coupled with various techniques such as fluorescence microscopy, surface plasmon resonance, and Raman spectroscopy.[29] For coupling with Raman spectroscopy, we implement a sphere-flat configuration (instead of the standard SFA crossed-cylinder geometry), and image analysis of Newton's Rings (NR) is used to measure the separation distance with sub-nanometer accuracy[29] rather than the multiple beam interferometry between back-silvered mica surfaces that is used in the standard SFA technique.

Here, we provide full design and testing details of the Raman-µSFA microscope, as shown in Figure 1. The microscope is designed to include independent light paths for illuminating and capturing NR images (blue), Raman excitation (green), and Raman signal collection (red), during approach of a sphere towards a flat coverslip within the µSFA. Crucially, the Raman excitation is focused at the center of the sphere-flat contact zone by a high NA objective, and NR images are acquired through the same objective. Therefore, we enable simultaneous acquisition of Raman spectra and the inter-surface distance, $D$. As a test case, we characterize fluoropolymer thin film interactions in water, exploiting two different modalities of the combined Raman-µSFA. We first discuss a *dynamic mode* in which we measure the force as a function of $D$ between the thin film and glass, and simultaneously obtain distance-dependent Raman spectra of the intervening water in the center of the contact region of the sphere-flat geometry. We also describe a *static mode* wherein we perform chemical-specific and label-free imaging of water and the fluoropolymer film with the surfaces in stable contact. The combined Raman-µSFA microscope provides a simple method for obtaining surface-sensitive spectroscopy and spatially resolved chemical imaging in all three dimensions: *x*- and *y*- via scanning microspectroscopy, and *z*- by correlating the signal strength with the separation distance.

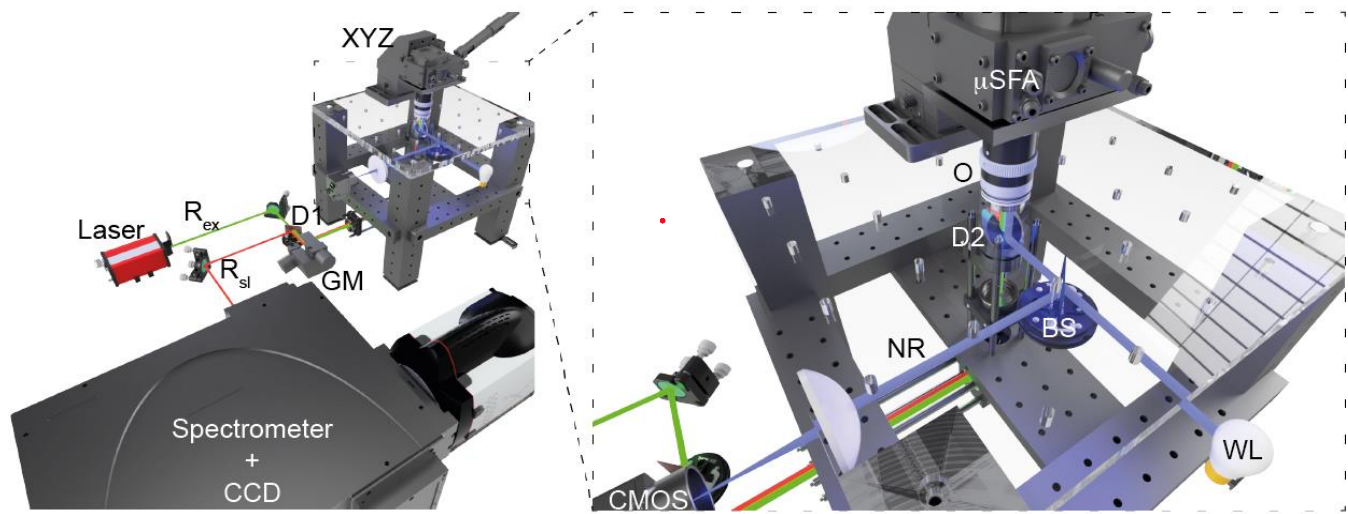

**Figure 1.** 3D rendering of the multimodal Raman-µSFA microscope. (left panel) Overview of the setup where the Raman imaging parts are highlighted: XYZ heavy-load coarse micrometer stage; the µSFA is mounted on the XYZ stage; laser for Raman excitation ($R_{ex}$); dichroic D1; GM, galvanometric mirror; Raman scattered light ($R_{sl}$). (right panel) A close-up view of the microscope stage, highlighting the light paths of the setup's three colors: green for $R_{ex}$, red for $R_{sl}$, and blue for the Newton's Rings (NR); D1, red/green dichroic mirror (532-nm shortpass); D2: blue/green/red dichroic mirror (490-nm longpass). O: objective. WL: white light source. BS: 50:50 beam-splitter. The top plate of the microscope stage, constructed from anodized aluminum, is here rendered transparent for clarity of presentation.

**Experimental**

The optical layout of the multimodal microscope is presented in Figure 1. A 2-dimensional diagram of the optical layout is provided in Supporting Information (Fig. S1) along with a detailed listing of all the optical components needed to build the setup. The Raman-µSFA microscope has been designed to perform simultaneous acquisition of Raman spectroscopy and NR imaging within the µSFA. The µSFA (SurForce LLC, *i*, see Supporting Information Table S1 for a list of the components corresponding to the following Roman numerals) is mounted on the confocal Raman microscope, which uses three color pathways without cross-talk between modalities: a 532-nm line for Raman microspectroscopy excitation (*ii*), a 600-nm-centered broadband line for spectroscopy detection, and a 488-nm-centered brightfield (*iii*) imaging in epi-configuration for the NR analysis.

The Raman confocal microscope is based on a focus raster scanning methodology (as opposed to sample raster scanning). Galvanometric mirrors (*iv*) are used for laser scanning to ensure mechanical stability of the SFA stage. The excitation light (*v*) is spectrally cleaned (*ii*), spatially expanded by a telescope, and steered onto a pair of galvanometric mirrors (*vi*). The midplane between the galvanometric mirrors are imaged, with a 4f telescope, on the back focal plane of a high numerical aperture (NA) objective (*vii*). The Raman inelastically backscattered light is collected by the same objective and follows the reverse path of the excitation light up to the galvanometric mirrors. In this way, the beam leaving the galvanometric mirrors does not move when coupled to a spectrometer, or a pinhole, upon raster-scanning at the image plane of the objective. The Raman light is then reflected by a short-pass filter (*viii*), spectrally cleaned with a notch filter (*ix*) and steered into the spectrometer (*x*) equipped with a charge-coupled device (CCD) camera (*xi*).

The source for the NR illumination is a white light lamp. The light is spectrally narrowed with an interference filter (*iii*), steered into the microscope with a long-pass dichroic mirror (*xii*), and focused (*xiii*) in the back focal plane of the objective. The contact plane is then imaged by the same objective via a beam splitter, a lens (*xiv*) and a complementary metal-oxide-semiconductor (CMOS) camera (*xv*). We magnify the image plane by 0.4, in order to correctly image at least two minima of the NR.

To illustrate the capabilities of the Raman-µSFA microscope, we performed force measurements and Raman spectroscopy on films of a commercial fluoropolymer poly[4,5-difluoro-

2,2-bis(trifluoromethyl)-1,3-dixole-co-tetrafluoro-ethylene], referred to here by its trade name Teflon-AF1600 (TAF, purchased from Sigma-Aldrich). The TAF is a copolymer of polytetrafluoroethylene (PTFE) and a fluorinated dioxole, the structure of which is shown in Figure 2A. The presence of the dioxole makes the polymer soluble in fluorinated solvents, and TAF is amorphous, yet exhibits many of the same physicochemical properties as PTFE. TAF films were prepared on glass coverslips by the following procedure. Coverslips were cleaned with piranha (3/1 by volume of $H_2SO_4/H_2O_2$), rinsed with copious amounts of Milli-Q water, and stored in Milli-Q water until further use. The glass was removed from Milli-Q, rinsed with absolute ethanol (Sigma-Aldrich), and blown dry with nitrogen. The clean and dry glass surface was then placed into a vacuum desiccator with a small vial of 1H,1H,2H,2H-perfluorodecyltriethoxysilane (FDTS, Sigma-Aldrich), and the desiccator was put under vacuum for ~30 minutes (~50 Torr). The desiccator was then placed in an oven at 60°C for 15 hrs. This procedure deposits a layer of FDTS on the glass surface, as evidenced by the high contact angle of a test water droplet (~100°). This FDTS acts as an adhesion layer for the spin coating of Teflon described in the next steps.

A solution of TAF was prepared by vigorous stirring of 1.5 wt% TAF in the fluorinated solvent Fluorinert FC-40 (Sigma-Aldrich) at 100 °C for 4 hrs. Using a micropipette, this solution was deposited onto the FDTS-functionalized glass and spin coated at 1000 rpm for 1 minute to deposit the TAF (Fig. 2A). After spin coating the surface was placed onto a 180°C hot plate for 1 minute to remove residual solvent and the surface was subsequently stored under vacuum until SFA experiments. The resulting TAF surface roughness and film thickness were measured by atomic force microscopy in dynamic mode (AFM, NanoSurf CoreAFM, Fig. 2B,C). The film thickness was determined by scoring the film with a scalpel and measuring the relative distance between the substrate and the film/air interface, as shown in Figure 2B. Figure 2C shows the vertically averaged line profile. The TAF film was furthermore found to be smooth, with root-mean-squared roughness of 0.3 nm, as shown in the insets of Figure 2C. Such a small roughness makes the TAF films particularly suitable for the SFA measurement at nanometric separations, despite slight inherent roughness on the underlying glass.

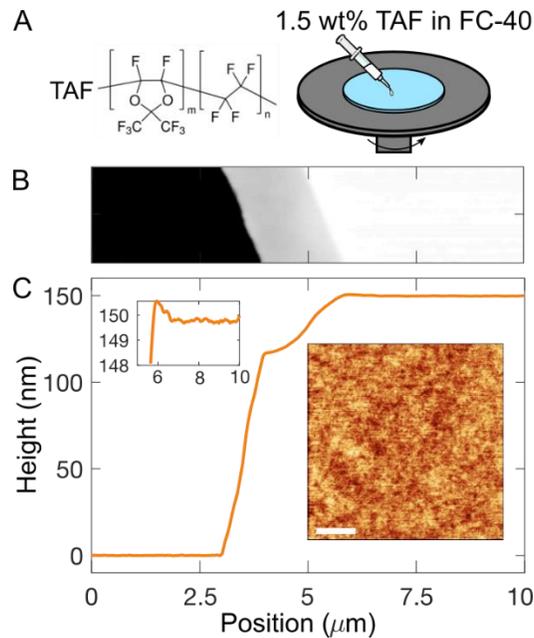

**Figure 2.** Teflon-AF (TAF) molecular structure and spin coating schematic for a TAF+FC 40 solution, forming a thin film. (B) AFM topography scan and (C) the vertically averaged line profile that results. Upper inset: a zoom of the upper film region with sub-nanometric roughness. Lower inset: AFM topography scan of a 10 × 10 μm² area on the film shown in (B). The horizontal scale bar is 2 μm and the color scale spans 2.3 nm.

For the SFA experiments, the TAF-coated coverslip was glued onto the bottom port of the μSFA. In the asymmetric glass vs. Teflon configuration, a freshly piranha-cleaned glass lens ($R \approx$ 2 cm, Edmund Optics #63476) was installed as the top surface. For the symmetric Teflon vs. Teflon configuration, a lens, coated with TAF following the same procedure as above, was installed as the top surface. Phosphate buffered saline (PBS, Sigma-Aldrich) solution was then injected between the surfaces. The standard configuration (glass vs. Teflon with PBS) is schematically shown in

Figure 3A. The curved surface approached the flat one while images of NR were recorded with the CMOS camera. Following the experiment, distances were measured using the NR analysis described recently[29] and summarized in Figure 3B. Briefly, we acquire images of the NR during the force measurement, which are radially-averaged around their symmetry point to obtain the radial intensity profile $I(r)$. This intensity profile is then used to reconstruct the surface shape, as is similarly done in reflection interference contrast microscopy (RICM).[30–32] The radial height profile $D(r)$ is calculated from $I(r)$ as shown in Equation 1,[32,33]

$$D(r) = \frac{\lambda}{4\pi n_1} \cos^{-1}\left(\frac{A-I(r)}{B}\right) , \qquad (1)$$

where $A = (I_{max} + I_{min})/2$, and $B = (I_{max} - I_{min})/2$. Equation 1 is valid within the first intensity extrema, and for parallel plate geometry.[32] Conveniently, TAF and water are nearly index matched ($n_{TAF} = 1.31$ and $n_{water} = 1.33$). We use $n_1 = 1.33$ for all of the measurements presented here. Parallel plates are a good approximation for the large radius $R \approx 2$ cm; the local slopes are $dD/dr \leq 5 \times 10^{-3}$ for all of the images presented here. For $r$ outside of the first intensity extrema, a height increment of $\lambda m/4n_1$ is used to calculate $D(r)$ at the extrema, where $m$ is the extrema number. A spherical model, corresponding to the lens shape, fits well to $D(r)$ using $R = 2.0$ cm, and the vertical offset provides $D(r = 0)$, the relevant distance for the force measurement.

The measured $I(r)$ when the surfaces are pushed into Teflon-glass contact indicates that the TAF film is about 140 nm in height (for $m = 1$ as discussed below) as indicated by the NR analysis of Figure 3B. Over several independent experimental setups, we find that the height of the film as determined by NR lies in the range of 130 nm to 160 nm. Figure 2C shows the height profile of an AFM measurement where the thickness is determined with nanometric precision giving a local height of 149 ± 1 nm, in quantitative agreement with the measurements by NR. The AFM measurement therefore gives an external calibration and confirms that $m = 1$ in the NR analysis. The Teflon-glass contact is here defined as $D = 0$ for the force measurement, $i.e.$, $D = D_i - D_0$, where $D_i$ is the measured distance at each point and $D_0$ is the contact value. $D$ therefore represents the thickness of the water film confined between the glass and the TAF (see Fig. 3A). As usual, the force is measured by spring deflection to obtain force-distance ($F$-$D$) measurements. Important aspects of the Raman microscope design, methods for correlating Raman spectra with the $F$-$D$ measurements, and chemical-specific imaging of confined contact zones are described in the following sections.

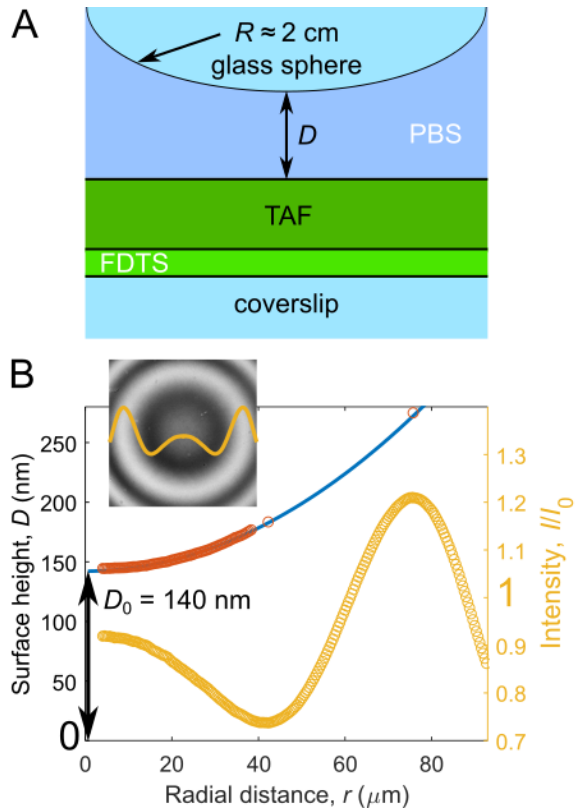

**Figure 3.** (A) Schematic of the layer structure in the µSFA for the glass-TAF setup (*n.b.* the scheme is not to scale). (B) Radial intensity profile $I(r)$ (yellow points, right axis) and a corresponding height profile $D(r)$ (red points, left axis) with a spherical fit (blue line), allowing one to obtain the distance offset, $D_0$. The inset shows the NR image for the corresponding intensity profile.

**Results and Discussion**

*Raman-µSFA design considerations*

We first describe important design aspects for achieving surface sensitivity with the spontaneous Raman effect. Although it has been demonstrated that Raman imaging has sub-monolayer-sensitivity in liquid-free environments,[34,35] the presence of surrounding bulk materials generate a signal which may overwhelm that of the intervening medium (see Sturm *et al.*[36] for a related discussion). Even though upon liquid confinement the bulk molecular vibrational response is suppressed, care must be taken to use optics that yield low-background signals from solid substrates (due to, *e.g.*, spurious fluorescence, luminescence, or Raman). We found that typical coatings (*e.g.* commercially available antireflection coatings) on glass surfaces generated too strong background, even rendering Raman measurement of bulk samples (when $D \geq 100$'s of nm) unfeasible. In our experience, bare glass surfaces yielded the lowest background levels in the spectral region >1400 cm$^{-1}$. Given that glass materials are so widely used, we consider this low-background result to be one of the principle advantages of our setup. Mica by contrast, often used in standard SFA, is birefringent and makes spectroscopic analysis more challenging for two reasons: (i) it distorts the point-spread function of the microscope,[37] lowering confocal capabilities (see below), and (ii) it scrambles the polarization state of the signal.[38] Functionalization of glass by silane chemistry is furthermore relatively straightforward. The glass surfaces are therefore an ideal choice for pairing µSFA with Raman spectroscopy. Standard background subtraction methods provided interfacial sensitivity in just a few seconds of integration time, in contrast to minutes in previous work.[24,26,27]

In addition to limiting the background signal, tight focusing is a second necessary aspect to increase surface sensitivity.[36] Given the approximate scaling of the interfacial signal *S*, and the bulk signal *B*, with the focus size $\omega_0$ in a confocal geometry ($S/B \sim 1/\omega_0^2$), reaching tight-focused, diffraction-limited performance can considerably increase sensitivity. However, we found that some of the optics used (dichroics, galvanometric mirrors) have low wavefront flatness therefore increasing the beam size slightly (*i.e.*, increasing the aberration of the wavefront). Despite the fact that we may not achieve diffraction limited performance, we show interfacial sensitivity below and anticipate that future improvements can be made by using adaptive optics.

The NR method for measuring distances is well suited for pairing with the tight focusing that is necessary for interfacial sensitivity in the Raman signal. The high NA objective nestles into the bottom port of the µSFA (see Fig. 1 and recent work[29] for more details), accommodating the high NA (1.4) and short WD (~100 µm) necessary for tight focusing. NR imaging is straightforward and provides ~0.5 nm distance resolution,[29] approaching the angstrom-level resolution of the multiple beam interferometry used in standard SFA. These design aspects allow for distance to be tightly correlated with the Raman signal as demonstrated below.

*TAF thin films*

Teflon is a technologically important polymer, yet the surface forces between Teflon surfaces have not been investigated in great detail, perhaps due to difficulties in surface preparation. Teflon surface forces have mostly been investigated in the context of low friction and van der Waals forces which are weak due to index matching in water or can even be repulsive in solvents such as cyclohexane.[39–42] Spin coating of TAF on a fluorinated glass surface (Fig. 2) provides a stable low energy surface, as exhibited by dewetting of water, ethanol, and even chloroform after several hours of immersion in the SFA.

TAF was chosen as a test case for the confocal Raman-µSFA for several reasons: it has a strong Raman signal, even at nanometer-level thicknesses, and exhibits a smooth interface (Fig. 2C). Here, we investigate the interactions between glass and TAF, as shown in Figure 3A. A spin coated TAF film with thickness ~140 nm mounted within the SFA has an easily resolvable Raman spectrum, as shown in Figure 4A. This spectrum is nearly identical to a previously published spectrum for TAF.[43] The small peak around 1000 cm$^{-1}$ is a result of the glass background signal, which is large in this region and therefore displays an artifact after background subtraction. Water between the TAF and glass exhibits a bulk-like Raman signal when the surfaces are far apart (water thickness $D = 180$ nm), as shown in Figure 4B. These measurements demonstrate that Raman spectra can be easily measured within the µSFA.

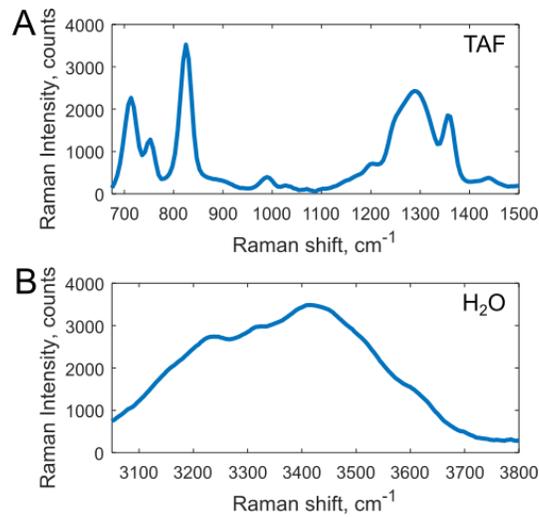

**Figure 4.** Raman spectra of the (A) TAF film and (B) the intervening bulk water were obtained within the µSFA when the water thickness was $D = 180$ nm.

*Confinement-dependent spectroscopy*

In the following, we establish and demonstrate two modes of operation for correlating separation distance with the Raman spectra of the confined contact zone. The dynamic mode allows simultaneous force-distance and spectroscopic measurements, providing powerful force-distance-spectroscopic correlations in real time. The static mode consists of bringing the surfaces into stable contact and correlating the static distance profile $D(r)$ with the Raman signal, allowing for longer signal averaging since $D(r)$ is constant when the surfaces are in contact.

*Dynamic mode: force-distance-spectroscopy correlation*

To acquire enough signal for interfacial sensitivity with the setup we described above, Raman spectra must be collected over a reasonable exposure time (~1 s). SFA measurements are often performed with slow approach/retract speeds of ~1-5 nm/s. The dynamic mode Raman-μSFA is thus performed by approaching the surfaces at these small velocities and acquiring Raman spectra synchronously with the NR images. As such, the Raman signal is effectively averaged over just a few nm, allowing for high sensitivity on both the SFA and Raman. When the surfaces encounter a repulsion, this speed effectively decreases and the Raman signal averaging occurs over much smaller distances. The laser is focused at the center of the NR, so the Raman spectra are measured at the distance of closest approach between the surfaces.

We tested the dynamic mode by measuring the interaction forces between TAF and glass immersed in PBS buffer, and correlating these interactions with Raman spectroscopic measurements of the intervening water. Figure 5 shows the procedure for dynamic correlation of the measured distance, force, and Raman signal. In this experiment, Raman spectra are taken synchronously with NR images at 1 frame per second while the surfaces are approached and separated at ~5 nm/s, and Figure 5A displays the measured spectra for the O-H stretch spectral region at different surface separations (the spectra in Fig. 5A have an offset due to a residual background). These spectra are averaged over the correlated times and distances, in which the colored spectra and numerical labels correspond with the colored and numbered bold lines in Figure 5B and 5C. Figure 5B shows the measured distance during the approach measurement and the integrated Raman signal (integrated over the spectral region of 3100-3650 cm$^{-1}$) measured simultaneously. The force $F = k\Delta D$ is then measured as the deviation $\Delta D$ from the constant slope in the $D$ vs. $t$ plot where $k$ is the spring constant, and the synchronized Raman signal can be correlated with the force at each point, as shown in Figure 5C. The O-H stretch intensity decreases linearly with decreasing separation distance, because both the vibrational density of states does not significantly change with distance, and the distance is always much smaller than the longitudinal extension of the 532-nm focus.

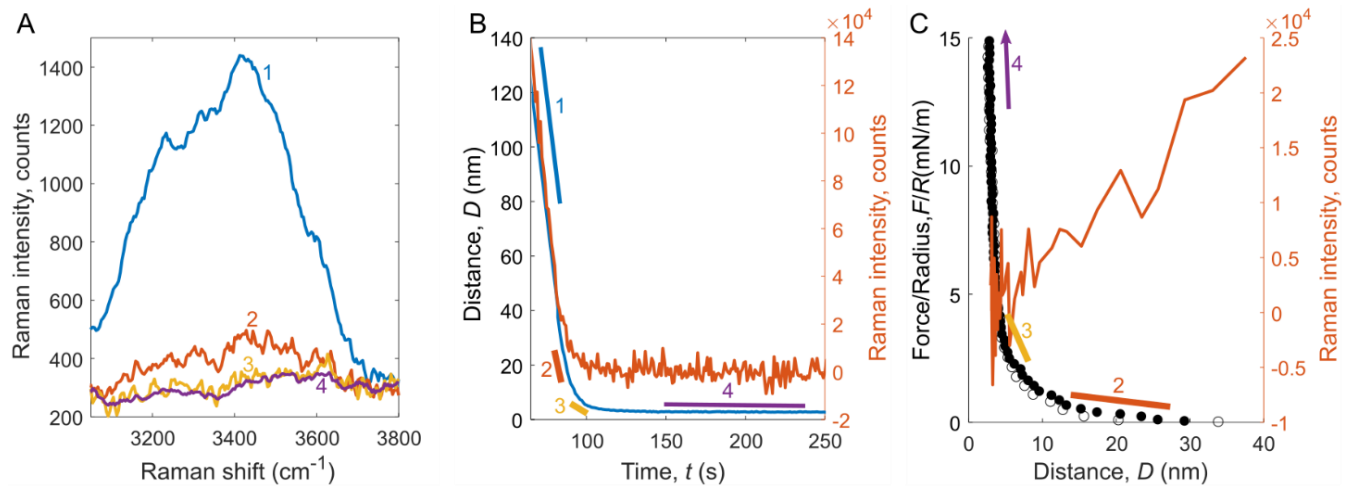

**Figure 5.** Correlating distance with Raman spectra in the µSFA. (A) Raman spectra of the O-H stretch are measured at each distance while the surfaces approach each other, and (B) the measured distance (left axis, blue curve) can be correlated with the integrated spectra (integrated over the spectral region 3100-3650 cm$^{-1}$, right axis, orange curve). The spectra in (A) are averaged over the times/distances indicated in (B), with the corresponding colors/numbers. The distance measured in (B) is used to calculate the (C) force vs. distance curve (left axis, black points: closed-approach and open-separation), and the correlated signal intensity is shown as the orange curve (right axis). The Raman signal and distance measurements are acquired synchronously with an acquisition rate of 1 frame per second.

The blue spectrum (1, Fig. 5A) is averaged from $135 > D > 80$ nm (Fig. 5B) and displays the bulk O-H stretch signal, similar to the bulk spectra shown in Figure 4B. This is in the linear regime of the $D$ vs. $t$ plot (Fig. 3B) where there is zero force. The red spectrum (2, Fig. 5A) is averaged from $26 > D > 13$ nm and shows intensity that is slightly blue-shifted (higher frequency).

The force is weak in this region (Fig. 5C) and significant water remains between the surfaces. The yellow spectrum (3, Fig. 5A) is averaged from $7 > D > 5$ nm and displays signal just slightly above that for the purple spectrum (4, Fig. 5A), which is averaged in the high force regime when the surfaces are stationary despite additional applied force (Fig. 5B and 5C). This slight decrease of signal for spectrum 4 compared to spectrum 3 is due to the last detectable water being squeezed out from between the surfaces.

At smaller inter-surface separations, *e.g.* in spectrum 3, we note a small peak of intensity around 3600 cm$^{-1}$. This wave number regime corresponds to the spectral region in which dangling hydroxide bonds (OH) can be inferred for water near hydrophobic interfaces.[44,45] While we cannot definitively claim evidence for such dangling OH bonds as a result of the residual background signal (which acts as an additional noise source), the intensity distribution does not exclude dangling OH in the TAF-water interfacial region; such an effect may be due to the strong hydrophobicity of the TAF surface. We find that the residual background noise, which remains even after the background subtraction routine, arises from the FDTS monolayer, and obscures the distance-dependent water signal.

Nonetheless, the changes measured in the Raman signal over the repulsive region of the force curve, *i.e.* $D < 30$ nm, are clearly measurable and indicate interfacial sensitivity in this region for the OH stretch signal with 1 second of integration time per point. Significant water signal remains at $D \approx 30$ nm, and the repulsion sets in around this distance (Fig. 3C). Over several independent experiments, this repulsive force was found to be approximately exponential with decay length between 7-9 nm. Electrostatic forces are vanishingly small at these distances (Debye length of PBS $\approx 0.8$ nm), so nanoscale roughness at the TAF surface and/or small asperities on the glass likely lead to this repulsion. In any case, the amount of detectable water vanishes to zero around $D \approx 5$ nm. This observation is notably different from our previous work with glass surfaces, for which a significant Raman signal in the contact region of glass-water-glass could be detected even at large compressive loads.[29] At this distance ($D \approx 5$ nm), the force increases steeply and is likely due to compression of the TAF film by the glass lens.

The combined Raman-µSFA approach can lead to unique insights that cannot be gained from either technique alone. For example, the local absolute distance at contact is difficult to measure accurately with the NR, especially when compared with the classic SFA technique, which directly gives the absolute distance. However, combining µSFA with the Raman allows for measurement of the "true" gap size at contact. Here, we benchmark this methodology with a measurement of glass vs. glass in water, because this system exhibits much weaker residual background noise compared to TAF films. We assume that the signal scales linearly with separation distance (as measured in Fig. 5C) such that the Raman signal intensity $I_o$ is some constant $C$ times the distance $D_o$,

$$I_o = CD_o \quad , \quad (2)$$

and the intensity after changing the distance by $\Delta x$ is

$$I_i = C(\Delta x_i + D_o) \quad . \quad (3)$$

Taking the ratio $R_{o/i} = I_o/I_i$ and rearranging, we have

$$\Delta x_i = D_o \left( \frac{1}{R_{o/i}} - 1 \right) \quad . \quad (4)$$

As such, we combine one measurement from each technique that has high confidence: the relative distance measurement from the NR (which is accurate to within 0.5 nm[29]) and the relative OH stretch integrated intensity from the Raman, allowing us to estimate the absolute separation distance at contact $D_o$. We performed a regression analysis using Equation 4 on the measured values of $\Delta x$ and $R_{o/i}$ over the full distance regime between glass-glass immersed in 10 mM KCl, as shown in Figure 6. The fitted value of $D_o = 6.1$ nm is slightly smaller than the measured value by NR of ~12 nm, but is likely closer to the actual value. Optical aberrations and slight increases in NR background intensity, as well as a slight mismatch in the optical indices of refraction, could lead to the slightly larger value from NR analysis. We attempted to perform this regression analysis on the TAF-glass system, but the level of residual background noise gave rise to physically unrealistic fitting parameters. Future work will focus on using different monolayer preparation to obtain smaller background signals.

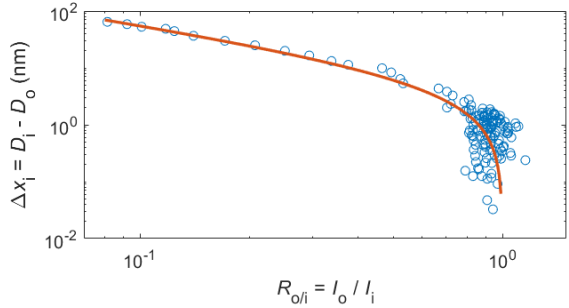

**Figure 6.** Regression analysis for a glass-glass system correlating the measured relative distance from NR analysis with the measured relative Raman intensity to find $D_o$ according to the analysis described in the text. Blue points are the data and the red line is the best fit value to Equation 4 with $D_o = 6.1$ nm.

It is instructive to estimate the limits of detection for the combined approach. Assuming a shot-noise-limited detection, the signal-to-noise (SNR) of the O-H stretch $I_w$ is given by

$$SNR \approx \frac{I_w}{\sqrt{I_B}} \qquad , \qquad (5)$$

where $I_B$ is the background intensity. Based on the accurate values measured for glass-glass at large distance, we obtain a photon flux of 5 photons/monolayer/s (assuming a 3 Å monolayer thickness), which leads to a 10 photons/s upon contact of the two hydrated surfaces. Based on the background photon flux we consistently obtained ($I_B=2\times10^3$ photons/s), a 30 s integration time would lead to SNR > 1, thus allowing to detect the spectrum of the two surface monolayers with finite acquisition time (note that this estimate considers a spectral pixel, not the integral as shown in Fig. 5B). This order of magnitude estimation shows that interfacial sensitivity is readily obtainable with the combined Raman-µSFA approach, opening the possibility for microspectroscopy of aqueous interfaces under confinement.

*Static-mode: label-free chemical imaging of confined TAF and water*

In static mode, a full range from bulk to confined Raman spectra is enabled by the SFA geometry: a single snapshot of the surfaces in sphere-flat contact provides distances from contact to bulk separations.[21] Thus, Raman signals can be acquired for much longer times while the surfaces are in stable contact, and distance-dependent spectroscopic insights can be obtained without moving the surfaces. Indeed, the interfacial sensitivity observed here is a result of the geometry of the confined sphere-flat contact, providing a well-defined contact area that is considerably larger than the laser spot.

To exploit the static mode, we brought the glass surface into contact with the TAF and performed chemically resolved imaging of TAF and confined water. Remarkably, we demonstrate chemical imaging with interfacial sensitivity of the confined liquid in just a few seconds of spectral acquisition. Figure 7 presents chemical maps of the contact region for the TAF (Fig. 7A) and water (Fig. 7B) vibrational response. The TAF image (Fig. 7A) is relatively featureless and shows that the TAF thin film is homogeneous and macroscopically flat over the contact area in the center of the image (~60 µm$^2$); a small slope over the field of view indicates that there is a small thickness change of the TAF film (~20%) that is common for spin-coated films over tens to hundreds of μm.[46]

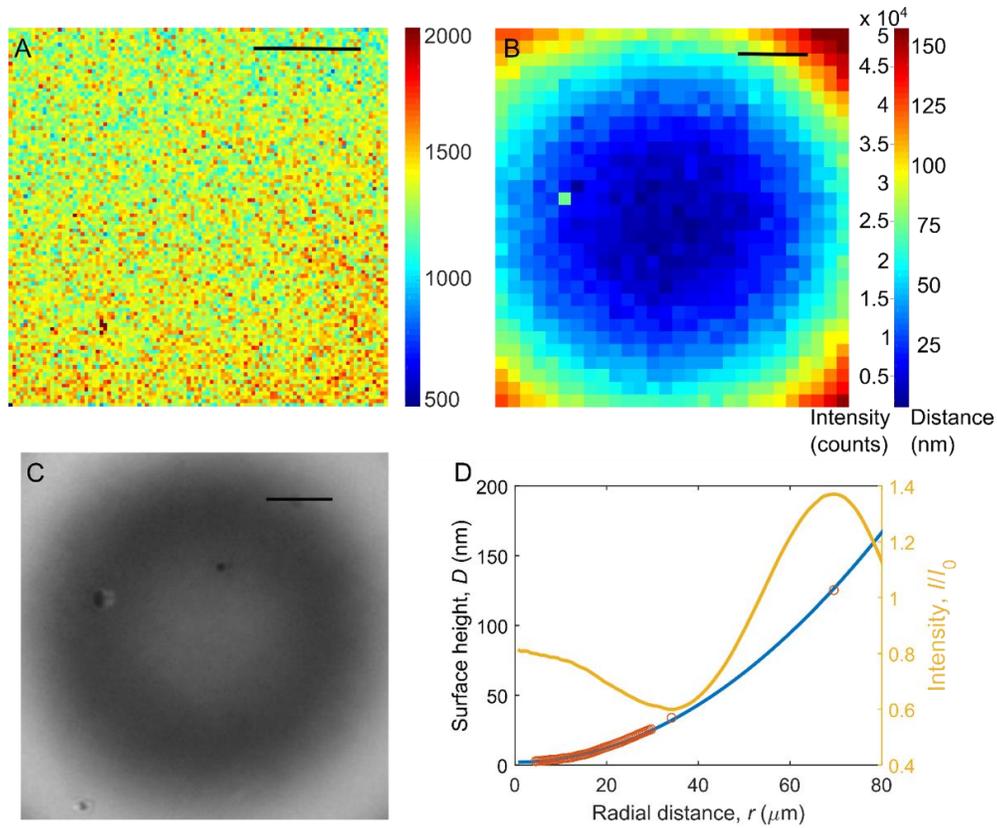

**Figure 7.** Chemical maps via Raman microspectroscopy within the μSFA of (A) TAF (signal integrated from 1220 to 1390 cm$^{-1}$, see Fig. 4A) and (B) water (signal integrated from 3100 to 3650 cm$^{-1}$, see Fig. 4B) during contact between TAF and glass in PBS. The O-H stretch intensity (left scale) can be mapped to the distance (right scale) as described in the text. The (C) bright field NR image corresponds with the water chemical map of (B). The (D) height profile (red points, left axis) is fitted to a spherical shape (blue line, left axis) and measured from the intensity profile (yellow line, right axis) generated from the NR bright field image. All scale bars = 20 μm. (A) was acquired with 50 ms pixel dwell time and (B) was acquired with 500 ms pixel dwell time.

Conversely, the O-H stretch integrated image (Fig. 7B) shows the expected spherical profile reminiscent of the lens shape. Figure 7C shows a bright field image that corresponds with the Raman signal of Figure 7B. By correlating the distance measured radially by the NR in Figure 7C, we can re-scale the intensity of Figure 7B to the separation distance. The correlated radial profile $D(r)$, corresponding spherical fit, and measured intensity profile $I(r)$ are shown in Figure 7D to a radial distance of $r = 80$ μm, which is approximately the maximum radial value displayed in Figures 7B and 7C. The intensity of the water signal scales linearly with $D$, as we showed in Figure 5C. Therefore the color intensity scale (left scale bar, Fig. 7B) can be directly mapped to the separation distance (right scale bar, Fig. 7B) according to $D(r)$ (Fig. 7D). These data indicate that a well-calibrated Raman signal can actually be used as a precise distance measurement, and the Raman intensity profile of water has been measured in all three spatial dimensions.

Finally, we examined the symmetric TAF-TAF configuration. Approaching the two TAF surfaces in degassed water results in a hydrophobic attraction between the two surfaces, which occurs from a large distance and results in nucleation of vapor at the contact zone. Such attractive forces and vapor nucleation were observed previously for fluorocarbon surfaces.[47] Vapor nucleates in the contact region and one can observe the surfaces jumping into contact, as well as water escaping from the hydrophobic contact. This attraction is likely due to nucleated vapor bubbles at defects in the TAF film, but the nucleation is too fast to observe the NR accurately. We attempted to obtain Raman microspectroscopy images of both the water and TAF after the dewetting process, but we observed laser damage in the film upon attempting to do so. No laser damage was observed over long exposure times in the Teflon-glass configuration discussed above.

Therefore, we examined the TAF-TAF configuration in air. First, we imaged a flat TAF surface with the TAF-coated sphere placed far away; similar to Figure 7A above, this image of TAF is relatively featureless, as shown in Supporting Information (Fig. S2). We then brought the two TAF into contact and separated them. Upon separation, we observed wrinkling of the film in the bright field microscope. We obtained a chemical-specific image of the film morphology, as shown in Figure 8. By comparing the chemical-specific image (Fig. 8, left panel) with the bright field image (Fig. 8, right panel), we observe that the TAF is clumped or lifted off the surface in areas of higher intensity and exhibits holes or depleted zones in areas of low intensity. Thus, we can induce microscopic morphological changes via contact and separation cycles, and observe such changes with chemical specificity, which greatly complements insights from bright field imaging. These examples of label-free chemical mapping point towards *in situ* chemical monitoring of physical and chemical surface morphology in interacting systems under confinement, including in tribology and wear studies, chemical-mechanical polishing, reactive or corrosive interfaces, and lipid domain morphology.

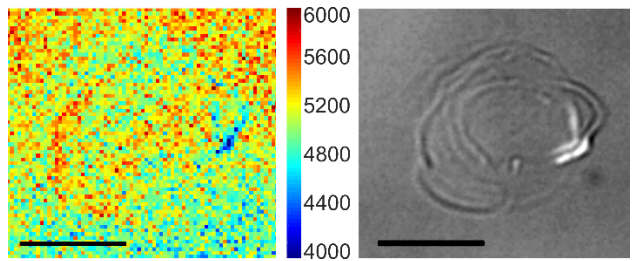

**Figure 8.** Chemical map of TAF, Raman signal intensity integrated from 1220 to 1390 $cm^{-1}$, after morphological changes due to contact with another TAF surface, which has been separated for taking the image. The chemical map (left panel) exhibits morphological features that correspond well with the bright field microscope image (right panel). The chemical image was acquired with 250 ms pixel dwell time, and the scale bar is 20 μm.

As designed here, the Raman and μSFA techniques are suitably compatible with each other and provide unique capabilities for simultaneous physical measurement and chemical identification. The simple geometry provides the opportunity to spectroscopically probe distances from contact to bulk by scanning a single line across the center of the contact. 3-dimensional chemical morphologies and the effects of confinement can be investigated. Chemical identification during confinement is an important advantage, for example in monitoring molecular species such as impurities or during interfacial chemical reactions. The simple optical setup used here is also a considerable practical advantage, compared to *e.g.* SFG, which requires expensive lasers and much more complex optical setups.

**Conclusions**

Here we presented a microscope and methodology for correlating surface forces studies with simultaneous *in situ* Raman spectroscopy. In dynamic mode Raman-μSFA, repulsive interactions between TAF films and glass across water were observed, and simultaneous Raman spectra reveal signal decreases for the water as the separation distance decreases. Signal averaging of the Raman shows slight modification of spectral features as the gap between TAF and glass decreases, potentially indicating that the population of water is modified in TAF-glass confinement, and these spectra correlate with the measured force. With static mode Raman-μSFA, chemical mapping of the TAF-water contact zone shows expected features: a mostly flat and featureless film for TAF, and an increase in signal with radial distance for the water, corresponding to the amount of water confined in the sphere-flat contact. Correlating this signal with the distance (*z* direction) and 2D imaging (*x-y*) allows for resolution of Raman spectra of confined geometries in 3D. Nanometer-level sensitivity is obtained for the water signal in less than 1 second of integration time. TAF-TAF interactions are attractive in water, resulting in vapor nucleation at the contact zone. A wear zone and resulting nanoscale features in the TAF film were chemically mapped.

The combined Raman-μSFA is a simple and relatively inexpensive method to obtain interface-sensitive spectroscopy and simultaneous insights into the physical interactions between surfaces. Indeed it is difficult to envision another combination of techniques that could provide similar multimodal insights. SFG has interface specificity and therefore interrogates purely the interface; there is no bulk signal and distance dependence would not be straightforward. The footprint of an AFM tip is too small to obtain a strong Raman signal dominated by the contact zone between the tip or colloid and a surface. Tip-enhanced Raman is limited to metallic contacts and provides a localized (surface) effect, and therefore cannot obtain long-range distance dependence. The current method will find broad uses in colloid and interface science, for example in correlating electrostatic double layer forces with hydrated ion spectroscopic signatures, observing structural or phase transitions of polymer brushes or melts upon applied force, chemical monitoring of corrosion under load, or observing lipid phase morphologies upon membrane fusion, to name a few.

**Supporting information:** 2D optical layout, list of optical components, movie of air nucleation between TAF-TAF in water, image of TAF in air.

## Acknowledgements

The authors were supported by LabEX ENS-ICFP: ANR-10-LABX-0010/ANR-10-IDEX-0001-02 PSL*. SHD acknowledges inspiration from the late Jacob N. Israelachvili, and this article is dedicated to his memory.

# Supporting Information

*Design details of Raman-µSFA*

We provide below more details on the setup and construction of the Raman-µSFA. A top view and side view of the optical layout is shown in Figure S1 to complement the 3-dimensional rendering shown in the text, and a list of the required optical components is shown in Table S1.

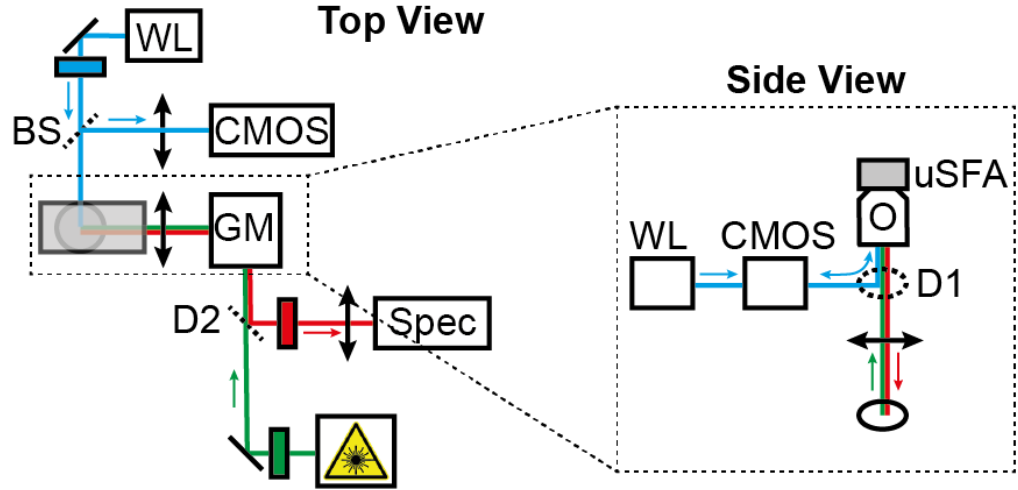

**Figure S1.** 2D optical layout of the Raman-µSFA. The abbreviations correspond with those in Figure 1 of the main text.

**Table S1.** List of optical components used in the construction of the Raman-uSFA

| item | description | manufacturer | specification |
|---|---|---|---|
| i | µSFA | SurForce LLC | |
| ii | optical filter; Raman excitation, 532/YY (center/width) | Thorlabs | FLH05532-4 |
| iii | optical filter; Newton's rings, 488 nm | Thorlabs | FL488-10 |
| iv | galvanometric mirrors | Thorlabs | GVS012/M |
| v | Raman excitation source, 532 nm | Oxxius | LCX-532 |
| vi | galvanometric mirrors | Thorlabs | GVS012/M |
| vii | objective, 60X, NA=1.4 (oil) | Nikon | |
| viii | short-pass filter | Thorlabs | DMSP550R |
| ix | notch filter | Thorlabs | NF533-17 |
| x | spectrometer | Andor | Shamrock 303i |
| xi | CCD camera | Andor | iXon Ultra |
| xii | long pass dichroic, 490 nm | Thorlabs | DMLP490 |
| xiii | lens, f = 150 mm | Thorlabs | AC254-150-AB-ML |

| xiv | lens, f = 80 mm | Thorlabs | AC508-080-AB-ML |
| xv | CMOS camera | Edmund | EO1312M |

*Chemical mapping of TAF in air*

Figure S2 shows a chemical map of the TAF film in air, before it was contacted with another TAF film. The film is observed to be uniform and featureless. After contacting with another TAF film and removing the contact, the morphology shown in Figure 8 of the main text was observed.

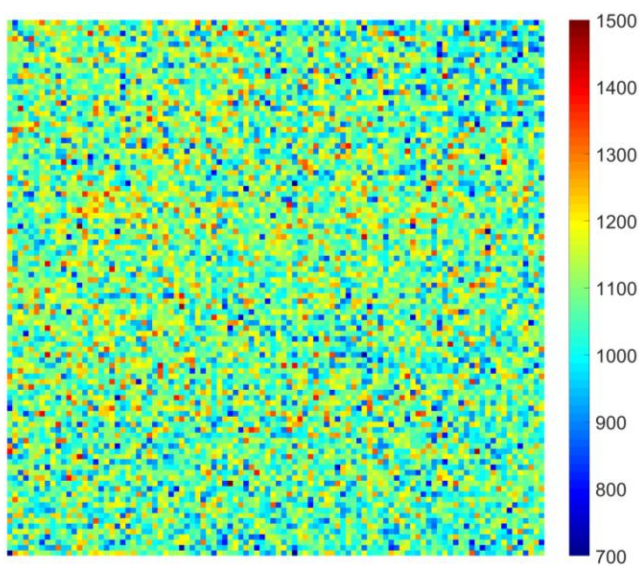

**Figure S2.** Chemical map of pristine TAF film in air. Color scale is the integrated intensity of the TAF spectrum from 1220-1390 cm$^{-1}$.